\documentclass[prl,a4paper,nofootinbib,
twocolumn,
]{revtex4-1}
\usepackage{graphicx}
\usepackage{bbold}
\usepackage{mathtools}
\usepackage{amsthm}
\usepackage{amsfonts}
\usepackage{amssymb}
\usepackage{slashed} 
\usepackage{color}
\usepackage{hyperref}
\usepackage{bm}
\usepackage{ulem}
\def\eq#1{{Eq.~(\ref{#1})}}

\def\tab#1{{tab.~(\ref{#1})}}

\def\etal{{\it et al.}}

\definecolor{oucrimsonred}{rgb}{0.6, 0.0, 0.0}
\definecolor{persianblue}{rgb}{0.11, 0.22, 0.73}
\definecolor{forestgreen}{rgb}{0.13,0.35,0.13}
 \hypersetup{colorlinks, citecolor=oucrimsonred, linkcolor=persianblue, urlcolor=oucrimsonred}
\def\hhref#1{\href{http://arxiv.org/abs/#1}{#1}} 
\newcommand{\be}{\begin{equation}}
\newcommand{\ee}{\end{equation}}
\newcommand{\bea}{\begin{eqnarray}}
\newcommand{\eea}{\end{eqnarray}}
\newcommand{\nn}{\nonumber}

\newcommand{\com}[1]{}
\newcommand{\gsim}{\lower.7ex\hbox{$\;\stackrel{\textstyle>}{\sim}\;$}}
\newcommand{\lsim}{\lower.7ex\hbox{$\;\stackrel{\textstyle<}{\sim}\;$}}
\begin{document}
\title[]{ Testing Bell inequalities  at the LHC with top-quark pairs
}
\date{\today}
\author{M.\ Fabbrichesi$^{\dag}$}
\author{R.\ Floreanini$^{\dag}$}
\author{G.\ Panizzo$^{\ast}$}
\affiliation{$^{\dag}$INFN, Sezione di Trieste, Via  Valerio 2, 34127 Trieste, Italy\\
$^{\ast}$Dipartimento Politecnico di Ingegneria ed Architettura, Universit\`{a} degli Studi di Udine, Via della Scienze 206, 33100 Udine, Italy  and \\INFN, Sezione di Trieste (Gruppo Collegato di Udine), via delle Scienze, 208, 33100 Udine, Italy}
\begin{abstract}
\noindent Correlations between the spins of top-quark pairs produced at a collider can be used to probe  quantum
entanglement at energies never explored so far. We  show how the  measurement of a single observable can provide a test of the violation of a Bell inequality  at the 98\% CL with the statistical uncertainty of the data already collected at the Large Hadron Collider, and at the 99.99\% CL with the higher luminosity of the next run. Detector acceptance, efficiency and migration effects are  taken into account. The test  relies   on the spin correlations alone and does not  require the determination of  probabilities---in contrast to  all other tests of  Bell inequalities.
\end{abstract}

\maketitle 
\textit{Introduction.---} A characteristic property of a quantum system is the  presence 
of quantum correlations (entanglement)   among its constituents not accounted for by classical physics (for a  review, see~\cite{Horodecki:2009zz}), leading to the violation of
specific constraints, the so-called Bell inequalities~\cite{Bell:1964kc,Bell2}.%
The violation of Bell inequalities requires the presence of the strongest version of quantum non-locality;
although weaker forms of non-classical correlations have been identified, 
they  play no role in our considerations.

Quantum correlations can readily be studied in a bipartite system
made of two spin-1/2 particles~\cite{Clauser:1978ng}.  This physical system is routinely produced at colliders
and the  spin correlations among quark pairs 
have  been shown~\cite{Bernreuther:2001rq,Bernreuther:2015yna} to be a powerful tool  in the physical analysis---limited aspects of which have been already studied by the experimental collaborations at the LHC on  data at 7 \cite{ATLAS:2012ao}, 8~\cite{Aad:2014mfk} and 13 TeV~\cite{CMS:2018jcg} of center-of-mass (CM) energy.

In this Letter, we focus on top-antitop pairs produced at the Large Hadron Collider (LHC) 
and identify a single observable  
probing the presence of quantum correlations among their spins. The measurement of such an observable
provides a test of a (generalized) Bell inequality.

Many experiments have been performed to test
analogous inequalities 
in various quantum systems involving photons and  atoms~\cite{Clauser:1969ny,Clauser:1974tg,Clauser:1978ng,Horodecki2,Genovese:2005nw}.
Similar tests  in the high-energy  regime of particle physics have been suggested  by means of $e^+e^-$ collisions~\cite{Tornqvist:1980af}, neutral meson physics~\cite{Benatti2,Banerjee:2016}, Positronium~\cite{Acin:2000cs}, Charmonium decays~\cite{Baranov:2008zzb}  and neutrino oscillations~\cite{Banerjee:2015mha}. No test has so far been performed at the high energies made available by the LHC---even though some  preliminary work has been done in \cite{Bernreuther:2001rq,Bernreuther:2015yna} and more recently in \cite{Afik:2020onf}. 
 In particular, we build on the results of \cite{Afik:2020onf} in which the entanglement of the top-quark pairs and the kinematical regions where it could be maximal were identified and  explicitly discussed.

Let us stress that all these tests involve the direct measurement of the joint
probabilities entering the various inequalities and therefore might be affected by the so-called loopholes, depending
on the specific characteristics of the used setups.
Our approach is quite different and unexplored: the focus is not on the probabilities of joint events, specifically 
top-quark pair spin projection measurements, but rather on their mutual spin correlations.  Such a measurement of correlations
 will provide evidence against a whole class of local
completions of quantum mechanics by explicitly exposing their internal
inconsistency. In order to be validated, these classical theories will need to reproduce both the probabilities
entering the Bell inequality and the averages of the spin correlation matrix through the presence of auxiliary stochastic variables and do that both at  atomic energies  and 
in the extreme relativistic setting of proton collisions at the LHC.

Reformulating the actual determination of the selected spin observable into a statistical test, we show how the value of this observable can be extracted from the events and the violation quantified at the confidence level (CL) of 98\% with the data already collected by the experimental collaborations at the LHC and   99.99\% CL (4$\sigma$ significance) with the higher luminosity of the next run. Detector acceptance, efficiency and migration effects have been  taken into account.

\vskip1em
\textit{Methods.---}  The quantum state of a two spin-1/2 pair, as the one formed by a top-quark pair system,
can  be expressed by the density matrix
%
\bea
\rho & =& \frac{1}{4}\Big[ \mathbb{1} \otimes \mathbb{1} + \sum_i A_i (\sigma_i\otimes \mathbb{1} )
+ \sum_j B_j(\mathbb{1} \otimes \sigma_j) \nn \\ 
& &+ \sum_{ij} C_{ij} (\sigma_i\otimes\sigma_j) \Big]\ ,
\label{rho}
\eea
%
where $\sigma_i$ are Pauli matrices, $\mathbb{1} $ is the unit $2\times 2$ matrix,
while the sums of the indices $i$, $j$ run over the labels 
representing any orthonormal reference frame in three-dimensions.
The real coefficients $A_i={\rm Tr}[\rho\, (\sigma_i\otimes {\bf 1})]$ and
$B_j={\rm Tr}[\rho\, ({\bf 1}\otimes\sigma_j)]$ represent the polarization
of the two spins, while the real matrix $C_{ij}={\rm Tr}[\rho\, (\sigma_i\otimes\sigma_j)]$
gives their correlations. 
In the case of the top-quark pair system, $A_i$, $B_j$ and $C_{ij}$ are functions of the parameters
describing the kinematics of the quark pair production.

In the CM reference frame of the top-quark pair system as produced at a $pp$ collider, the two spin-1/2 quarks  fly apart
in opposite directions.
One can then extract  the probability $\mathcal{P}(\uparrow_{\hat n}; -)$
of finding the spin of one quark in the state $\uparrow_{\hat n}$, with the projection of the spin along
the axis determined by the unit vector $\hat n$ pointing in the up direction.
Similarly, one can  consider double probabilities, like $\mathcal{P}(\uparrow_{\hat n}; \downarrow_{\hat m})$,
 of finding the projection of the spin of the quark along the unit vector $\hat n$
pointing in the up state, while the companion antiquark has the projection of its spin 
along the direction of a different unit vector
$\hat m$ pointing in the down state.

In  classical physics, 
these probabilities involve averages over 
suitable distributions of  variables and obey the following  (generalized) Bell inequality~\cite{Clauser:1974tg}: 
\bea
& &\mathcal{P}(\uparrow_{{\hat n}_1}; \uparrow_{{\hat n}_2}) - \mathcal{P}(\uparrow_{{\hat n}_1}; \uparrow_{{\hat n}_4})
+\mathcal{P}(\uparrow_{{\hat n}_3}; \uparrow_{{\hat n}_2}) + \mathcal{P}(\uparrow_{{\hat n}_3}; \uparrow_{{\hat n}_4})
\nn \\
& & \leq \mathcal{P}(\uparrow_{{\hat n}_3}; -) + \mathcal{P}(-; \uparrow_{{\hat n}_2})\ ,
\label{Bell-inequality}
\eea
where ${\hat n}_1$, ${\hat n}_2$, ${\hat n}_3$ and ${\hat n}_4$
are four different three-dimensional unit vectors determining four spatial directions
along which the spins of the quark and antiquark can be measured.
In quantum mechanics the same probabilities are 
computed as expectation of suitable spin-observable
operators in the state (\ref{rho}), so that the previous inequality reduces to the following constraint
\begin{equation}
\Big|{\hat n}_1\cdot C \cdot \big({\hat n}_2 - {\hat n}_4 \big) +
{\hat n}_3\cdot C \cdot \big({\hat n}_2 + {\hat n}_4 \big)\Big|\leq 2\ ,
\label{algebraic-inequality}
\end{equation}
involving only the spin correlation matrix $C_{ij}$ and not the polarization
coefficients $A_i$ and $B_j$.

In order to test the Bell inequality in \eq{algebraic-inequality}, one needs to experimentally determine the matrix $C$
and then suitably choose  four spatial directions ${\hat n}_1$, ${\hat n}_2$, ${\hat n}_3$ and ${\hat n}_4$
that maximize the left-hand side of (\ref{algebraic-inequality}).
In practice, there is no need to optimize the choice of ${\hat n}_i$: this maximization process has already been performed in full generality in Ref.~\cite{Horodecki2},
for a generic spin correlation matrix.
Indeed, consider the matrix $C$ and its transpose $C^T$ and form the symmetric, positive, $3\times 3$ matrix
$M= C^T C$ whose three eigenvalues $m_1$, $m_2$, $m_3$ can be ordered by decreasing magnitude:
$m_1\geq m_2\geq m_3$. 
The two-spin state density matrix $\rho$ in (\ref{rho}) violates the inequality (\ref{algebraic-inequality}), or
equivalently (\ref{Bell-inequality}),
if and only if the sum of the two greatest eigenvalues of $M$ is strictly larger than 1, that is
\begin{equation}
m_1 + m_2 >1\, .
\label{eigenvalue-inequality}
\end{equation}
In other words, given a spin correlation matrix $C$ of the state $\rho$ that satisfies (\ref{eigenvalue-inequality}),
 there are for sure choices for the vectors ${\hat n}_1$, ${\hat n}_2$, ${\hat n}_3$, ${\hat n}_4$
for which the left-hand side of (\ref{algebraic-inequality}) is larger than 2. 

It should be stressed that the above formulation, based on the relation (\ref{Bell-inequality}),
departs from the more standard approaches adopted in testing Bell inequalities, in
particular in quantum optics. While in the standard, direct tests one needs to
experimentally determine the expectation values of spin observables entering the Bell inequalities,
in the above, indirect approach the actual measure of probabilities is avoided, in favor of
the determination of the spin correlation matrix $C$---the  entries of which can be 
measured by studying the kinematics of the quark-antiquark decay products~\cite{Bernreuther:2015yna}.

In the recent analysis~\cite{CMS:2018jcg}, the spin correlations of the top-quark pairs produced at the LHC are analyzed but  only after  being averaged over
a large portion of phase space; the  values obtained for the entries of $C$  are
 in agreement with the inequality (\ref{algebraic-inequality}), for any choice of the four vectors ${\hat n}_i$.
This agreement is the consequence of the averaging procedure (mixing) that unavoidably reduces the
entanglement content of the density matrix $\rho$.

On the other hand, the  study in~\cite{Afik:2020onf} 
 suggests that by focusing on specific, small regions
of the  phase space, the entanglement of the top-quark pair state
could be close to maximal (see also~\cite{Cervera-Lierta:2017tdt}) and the Bell inequality in \eq{Bell-inequality} could be violated at the maximal level.

\vskip1em
\textit{Results.---} The sum of the eigenvalues $m_1+m_2$ provides an observable whose value, as extracted from the data, tests whether the Bell  inequality in \eq{algebraic-inequality} is violated or not. To compute this observable we collect all the entries of the correlation matrix $C$ as given in the process
\be
pp \to t+\bar t  \to \ell^\pm \ell^\mp  + \text{jets} + E_T^{\text{miss}}\, ,
\label{eq:process}
\ee
where $\ell = e$, $\mu$ are taken only in different-flavor combinations, in order to better connect with experimental measurements in this channel.  $E_T^{\text{miss}}$ stands for the transverse missing energy.

We simulate full matrix elements for the top quark production and decays through the decay chain formalism built into \textsc{MadGraph5}~\cite{MG5}, which embeds full spin correlations and Breit-Wigner effects, thus excluding only non-resonant  
diagrams. Within the Standard Model we consider gluon ($gg$) and quark ($q\bar q$) initiated top-quark pair production at leading order in the strong and electroweak couplings, using the NNPDF23 \cite{Ball:2012cx} leading order parton distributions set and within the four flavor number scheme, thus fully taking into account for bottom quark mass effects. Next-to-leading order corrections in the strong coupling are known to be small on the largest entries of $C$: 
at the LHC energies their impact  is less than 2\%~\cite{Bernreuther:2001rq}, and will be neglected in the following. 
We assume a CM energy of 13 TeV, setting both the renormalization and factorization scales to the sum of the transverse energies of the final state particles.

\begin{figure}[h!]
\begin{center}
\includegraphics[width=3in]{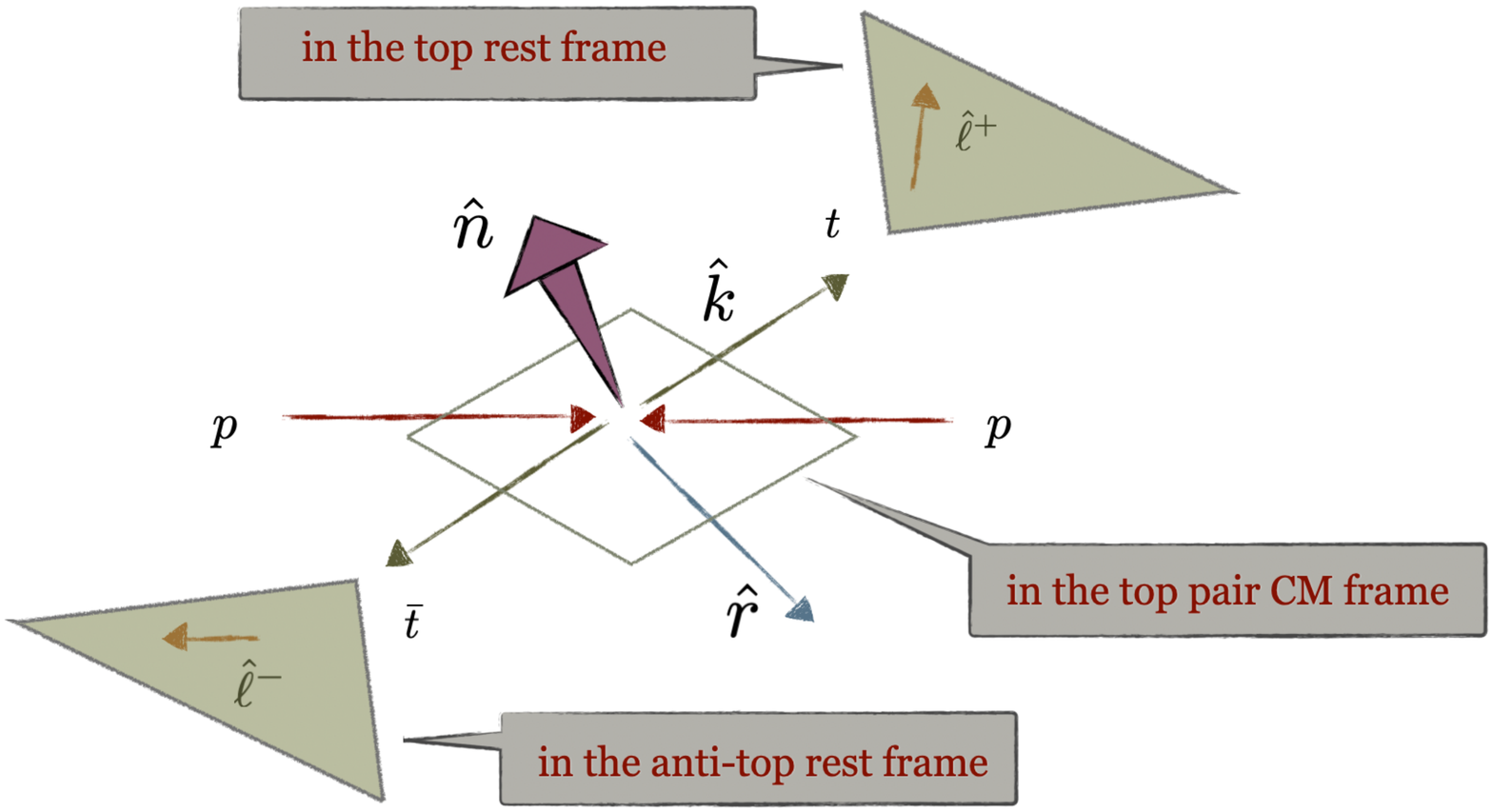}
\caption{\small Kinematics and coordinate systems used in the analysis. The $t$ and $\bar{t}$ rest frames are reached from the $t\bar{t}$ center of mass frame by rotation-free boosts. 
\label{fig:coordinates} 
}
\end{center}
\end{figure}

 We  follow~\cite{Bernreuther:2015yna} for the choice of orthonormal basis for the matrix $C$ of \eq{rho}. The unit vectors $\hat{r}$ and $\hat{n}$ are built  starting from the direction of flight $\hat{k}$ of the top quark  in the top pair CM frame with respect to one of the proton beams directions in the laboratory frame $\hat{p}$:
\be
\hat{{p}} = (0,0,1),~~\hat{r}=\dfrac{1}{r}(\hat{p}-y\hat{k}),~~\hat{n}= \dfrac{1}{r}(\hat{p}\times \hat{k})\, ,
\label{eq:axes}
\ee
(see Fig.~\ref{fig:coordinates}) where
\be
y=\hat{p}\cdot \hat{k}=\cos\Theta,\quad r=\sqrt{1-y^2},
\ee
  and $\Theta$ represents the top-quark scattering angle.
The correlation matrix $C$ can  be experimentally accessed through the angular spin correlations of the $t\bar{t}$ leptonic decays---whose directions of flight in the $t$ and $\bar{t}$ rest frames  are described, respectively, by the unit vectors $\hat{\ell}_{\pm}$. 
 These angular spin correlations are determined by properly averaging the products $\xi_{ab} = \cos \theta^a_+ \cos \theta^b_-$, where we defined the quantities 
\be
\cos \theta^a_+ = \hat{\ell}_+\cdot \hat{a}\quad \text{and} \quad \cos \theta^b_- = \hat{\ell}_-\cdot \hat{b}\, ,
\ee
and the labels $a$ and $b \in \{ \text{k}, \text{n}, \text{r}\}$ follow the conventions of \tab{tab:axes} for the choices of reference axes. Indeed 
one can show that, in the absence of acceptance cuts, the elements of the 3$\times$3 matrix $C$ can be expressed as \cite{Bernreuther:2015yna}
\be
C_{ab} \big[ \sigma (m_{t\bar{t}}, \cos \Theta) \big] = - 9 ~\dfrac{1}{\sigma} \int d\xi_{ab}  \dfrac{d \sigma}{d \xi_{ab}} \xi_{ab}\, ,
\label{eq:Cexp}
\ee
 with the residual dependence of the cross section $\sigma$ on $\cos \Theta$ and the invariant mass $m_{t\bar{t}}$ of the top-quark pair system being understood. The integral of \eq{eq:Cexp} represents precisely the average of the products $\xi_{ab}$ taken over the leptonic angular phase space. 

\begin{table}
\begin{tabular}{c | c c }
  ~$a$, $b$~ & $\hat{a}$ & $\hat{b}$ \\ \hline
 ~k~ & ~$\hat{k}~$ & ~$-\hat{k}$ ~\\
 ~n~ & ~sgn($y$)$\hat{n}$ ~& ~-sgn($y$)$\hat{n}$ ~\\
 ~r~ &  ~sgn($y$)$\hat{r}$ ~& ~-sgn($y$)$\hat{r}$ ~\\
\end{tabular}
\caption{Notation for the labels $a$, $b$ and corresponding choice of reference axes, following the definitions of \eq{eq:axes}} \label{tab:axes}
\end{table}

In order to fully take advantage of \eq{eq:Cexp}, the event generation was performed removing any possible kinematic cuts, both in production and decay. The diagonalization of the matrix $C$, needed to test \eq{eigenvalue-inequality}, can  be performed as a function of $m_{t\bar{t}}$ and $\Theta$. 

\begin{figure}[h!]
\begin{center}
\includegraphics[width=3.4in]{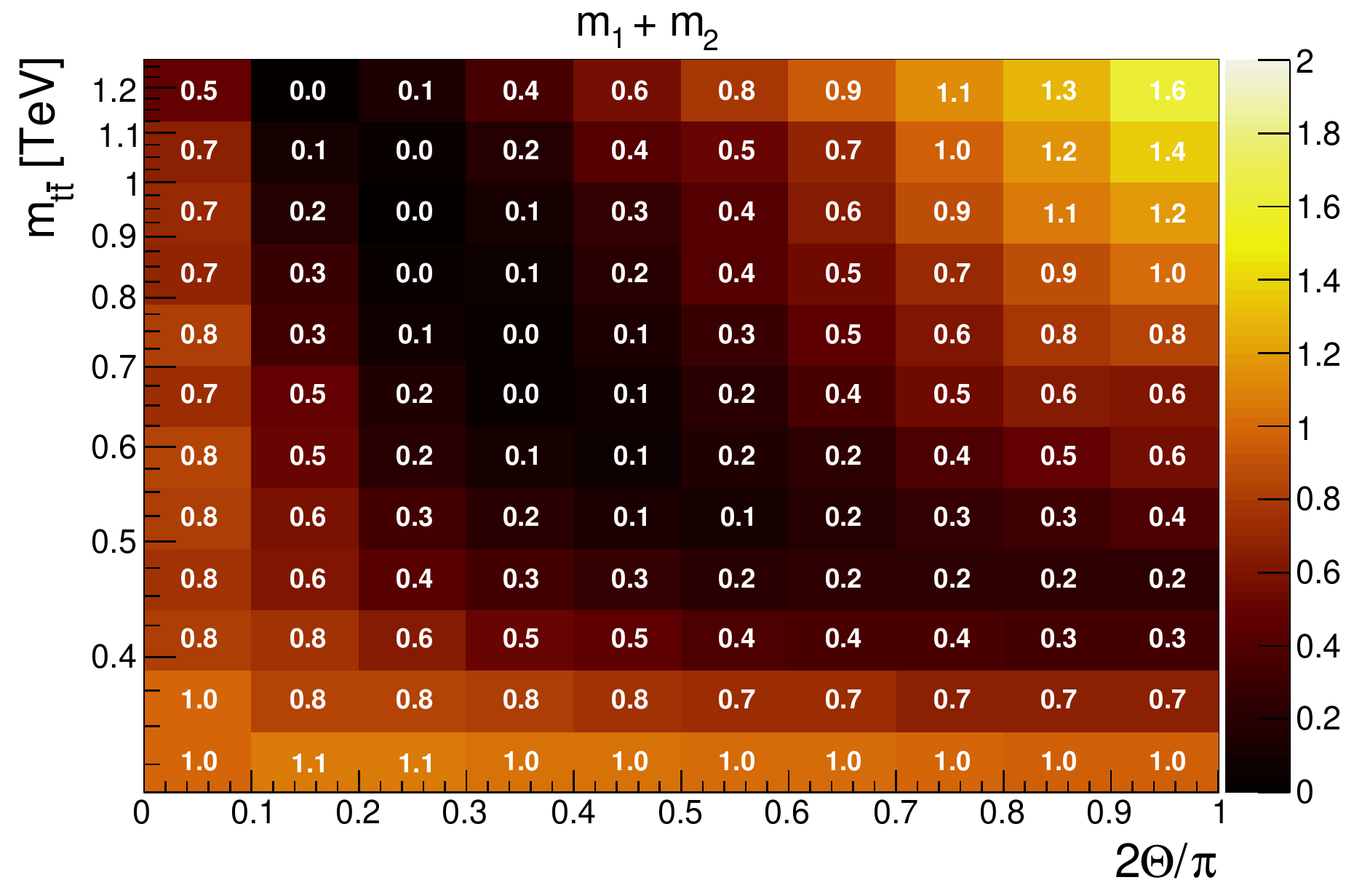}
\caption{\small Values of the observable $m_1+m_2$ in the phase space of the invariant mass $m_{t\bar t}$ vs.\ the scattering angle $\Theta$. The last bins in $m_{t\bar t}$ include overflow events. Bins in the upper right corner have the largest values of $m_1+m_2$ and are selected for testing  the violation of the Bell inequality.
\label{fig:m1m2} 
}
\end{center}
\end{figure}

The result of this procedure is shown in Fig.~\ref{fig:m1m2}, whose event statistics benefits from the intrinsic initial-state symmetry $\Theta \rightarrow \pi - \Theta$. The binning choice represents the best compromise between the expected event statistics at the LHC and the unavoidable dilution of entanglement effects coming from averaging $\xi$ in bigger portions of phase space. 

We can identify  in Fig.~\ref{fig:m1m2} two regions where \eqref{eigenvalue-inequality} holds, one at $m_{t\bar{t}}$ close to threshold, and another at $m_{t\bar{t}} \gtrsim 0.9$ TeV and $2 \Theta/\pi \gtrsim 0.7$.  Of these two regions, only the one at large $m_{t\bar{t}}$ presents a constructive sum of the   $q\bar{q}$ and $gg$ contributions, both giving rise to a top-quark pair state close to a pure, maximally
entangled state \cite{Bernreuther:2015yna,Afik:2020onf}, and therefore  increase $m_1+m_2$.  In the other region close to threshold, $q\bar{q}$ events produce a mixed state and dilute the $gg$ pure maximally entangled state---even if the $q\bar{q}$ contribution is subdominant in terms of cross section rates. This difference explains the higher values of $m_1+m_2$ in the top right corner of Fig.~\ref{fig:m1m2}, where $m_1+m_2$ is expected to reach a value as large as $1.6$.

In order to assess to what extent   the values of \hbox{$m_1+m_2>1$}  can realistically be used to prove a violation of the generalized Bell inequalities \eq{algebraic-inequality}, we study the impact on their determination of statistical uncertainties due to detector resolution, acceptance, efficiency and migration effects. To this aim, we reformulate the problem into a statistical test of the null hypothesis $\{H_0:\, m_1+m_2\leq 1\}$. We  compute the significance of the corresponding outcome at the LHC with 139 fb$^{-1}$, the full Run II luminosity, by performing 10 new, independent simulations of the process in \eq{eq:process}, with fast simulation of the ATLAS detector using the {\sc Delphes} \cite{deFavereau:2013fsa} framework. 

We require  at least two anti-$k_t$ jets with $R=0.4$ and at least one $b$-tagged jet, all with transverse momentum $p_T>25$ GeV and rapidity $|\eta|<2.5$. Similarly, both $e^\pm$ and $\mu^\pm$ leptons are required to have $p_T>20$ GeV and $|\eta|<2.47$. The neutrino momenta from the dileptonic decay are not directly detectable, since only their sum can be inferred through the missing transverse energy $E^T_\text{miss}$ of the event. The $t$ and $\bar{t}$ momenta need thus to be reconstructed using the neutrino weighting technique~\cite{Abbott:1997fv}. With this method, the sums of the momenta of the candidate reconstructed neutrinos, charged leptons and  candidate $b$ jets are constrained to satisfy four equations on the invariant mass of the two candidate $W$ bosons and top quarks. The possible solutions of this unconstrained system are assigned a weight $w$.
The solution which maximizes $w$ is eventually used to reconstruct the $t$ and $\bar{t}$ momenta for that event.
 This procedure allows us to determine the reconstructed 
distributions which are eventually corrected for detector resolution and acceptance
effects using a simplified unfolding procedure. 
The good agreement between our response matrix and those published for comparable processes~\cite{ATLAS:2019zrq} shows that migration effects have been properly simulated.

Tuning such pseudo-experiments to have a statistics equal to the one expected with present LHC luminosity, we can take the resulting standard deviations $s_i$ on \hbox{$m_1+m_2$} as the predicted statistical uncertainty,  with detector effects included.  
 In testing the hypothesis, we use a standard $\chi^2$ statistical test,
\be
\chi^2 = \sum_i \dfrac{(1-m^i_1-m^i_2)^2}{s_i^2}\, ,
\ee
where the sum runs over the set of bins that maximize the Standard Model expected significance for $m_1+m_2>1$. 

We find that, under such conditions, the null hypothesis and the violation of \eq{algebraic-inequality} can be assessed at the 98\% CL with present Run II luminosity. Moreover, after rescaling this result by the projected luminosity of the LHC full Run III, we expect that it will be possible to test the violation  at the  99.99\%  CL ($4\sigma$ significance). 

 While systematic uncertainties associated with the unfolding procedure itself are known to be negligible \cite{ATLAS:2019zrq}, the  results hereby presented do not include other theoretical and experimental systematic effects, 
the inclusion of which is  beyond the scope of the present Letters since it would require a more detailed simulation both of the detector and of the collisions conditions, which is only possible   for the experimental collaborations. 

\vskip1em
\textit{Conclusions.---}  We have shown that the measurement of a single, suitable defined observable of the top-quark  pair system can be used to  ascertain quantum correlations among the spins of two quarks and in turn  test a Bell inequality 
with the data already collected at the LHC. 
This test can  provide clear evidence for quantum mechanics in an energy range never explored before.

Testing Bell inequalities at high-energy colliders differs from
 the more familiar tests performed using quantum optics experiments. 
In the latter, the request of Bell locality
can be easily achieved by using spin polarization measurements that are space-like separated;
these measurements cannot influence each other, so that the two events have independent statistics.
It is not possible to follow the same procedure in the case of the top-quark system because of the obvious restrictions of the detectors employed at the LHC.

Nevertheless, there are advantages in studying Bell's inequalities in a high-energy settings along the lines
discussed in this Letter.
In quantum optics tests, to avoid so-called loop-holes connected to the lack of control of the
number of pairs produced that actually impinge in the detectors, and other inevitable inefficiencies,
one is generally forced to use inequalities more involved than (\ref{Bell-inequality}).
Although loop-hole free test of Bell inequalities have been recently performed
in quantum optics \cite{Hensen:2015ccp,Giustina}, 
none of these problems affect the (indirect) Bell test presented above, as it reduces to the study of
the spin correlation matrix $C$, without the need of any {\it a priori} commitment about efficiencies of detectors.

 We believe that our results will stimulate additional analyses to test Bell inequalities by means of the actual experimental data
collected at LHC and motivate further investigations on other possible tests of quantum mechanics
at high-energy colliders.

{\small\textit{Acknowledgements.---} We thank M.\ Pinamonti for help on the data analysis.
MF is affiliated to the Physics Department of the University of Trieste,  the Scuola Internazionale Superiore di Studi Avanzati
and the Institute for Fundamental Physics of the Universe, Trieste, Italy---the support of which is gratefully acknowledged. RF acknowledges that his research was conducted within the framework of the Trieste Institute for Theoretical Quantum Technologies.}

\vskip -.5cm



\end{document}